# Structural properties and spin-phonon coupling in orthorhombic Y-substituted GdMnO$_3$

R. Vilarinho[1], E. C. Queirós[2], A. Almeida[1], P. B. Tavares[2], J. Agostinho Moreira[1]

1-IFIMUP and IN-Institute of Nanoscience and Nanotechnology, Departamento de Física e Astronomia da Faculdade de Ciências, Universidade do Porto, Rua do Campo Alegre, 687, 4169-007 Porto, Portugal.

2-Centro de Química - Vila Real, Departamento de Química. Universidade de Trás-os-Montes e Alto Douro, 5000-801 Vila Real, Portugal.

## Abstract

We present a systematic study of the structure and lattice dynamics at room temperature, and the phonon behavior at low temperatures in orthorhombic Gd$_{1-x}$Y$_x$MnO$_3$ manganites, with $0 \leq x \leq 0.4$, using powder x-ray diffraction and Raman scattering. A thorough analysis towards the correlation between both structural and Raman modes parameters was undertaken. The data obtained at room temperature reveal structural distortions arising from the Jahn-Teller distortion and octahedra tilting. The Jahn-Teller distortion is apparently x-independent, while an increase of Mn-O1-Mn bond angle of about 0.5° could be ascertain when *x* changes from 0 to 0.4. Spin-phonon coupling was evidenced from the Raman results. The temperature dependence of the B$_{1g}$ in-plane O2 stretching mode of the MnO$_6$ octahedron has revealed either a positive or negative shift regarding the pure anharmonic temperature dependence of the phonon frequency, which strongly depends on the Y-concentration. The frequency renormalization is explained in terms of a competition between ferro and antiferromagnetic interactions. The ratio between the spin-phonon coupling constant and the effective magnetic exchange integral per spin was determined from the renormalized wave number of the in-plane O2 stretching mode, associated with the spin-spin correlation function.

# 1. Introduction

Magnetically-induced ferroelectricity is an important issue of to date research, because the control of polarization through electric/magnetic fields opens new insights into fundamental knowledge, and assumes a central role for new devices, namely for electric-control of the spin transport. Ferroelectricity is a ground state characterized by the emergence of a spontaneous polarization, which is evidently associated with inversion center breaking lattice distortions, stemming from charge displacement or existent dipoles reorientation. Consequently, the mechanisms driving magnetically induced ferroelectricity undoubtedly involve the interplay between spins and lattice. The understanding of the mechanisms underlying the emergence of the magnetoelectric effect in single phase materials is highly requested for the design of new structures which enhance multiferroic properties.

Among the compounds exhibiting magnetically-induced ferroelectric phases, the metal transition oxides are the most attractive, as they can be chemically modified in order to tune their physical properties, namely the magnetoelectric effect. Among them, the perovskite rare-earth manganites $R$MnO$_3$, $R$ being a rare-earth ion, occupy a field of special focus. In these magnetoelectric materials, the spin-phonon coupling and magnetically-induced ferroelectricity have been experimentally evidenced. Due to these properties, a direct action of applied electric (magnetic) field on the spin structure (lattice) is expected. Moreover, in the rare-earth magnetoelectric materials, ferroelectricity is of improper nature, arising in modulated magnetic phases, according to the Dzyaloshinskii-Moriya interaction. It is well stablished that orthorhombic $R$MnO$_3$ compounds exhibit competitive ferro and antiferromagnetic interactions, whose balance can be tuned by structural deformations promoted by either controlled atomic substitution and uniaxial strain in thin films.[1–3] The fine-tuned balance between competitive ferro and antiferromagnetic interactions stabilizes the suitable modulated magnetic phases allowing for ferroelectricity. No wonder why these materials are very much studied in both experimental and theoretical points of view. In this framework, the Gd$_{1-x}$Y$_x$MnO$_3$ system, with $0 \leq x \leq 0.4$, is an interesting case of study.

The ($x$, T) phase diagram of Gd$_{1-x}$Y$_x$MnO$_3$ system, with $0 \leq x \leq 0.4$, reflects the effect of lattice distortions induced by the substitution of Gd$^{3+}$ by smaller Y$^{3+}$ ions, which gradually unbalances the antiferro against the ferromagnetic exchange interactions, enabling the emergence of ferroelectricity for higher enough concentrations of yttrium ion.[4] The proposed phase diagram is presented in Ref. [4]. The compounds belonging to this system undergo a transition from a paramagnetic to a sinusoidal incommensurate antiferromagnetic phase, at T$_N \approx$ 42 K. For $x \leq 0.1$, the paramagnetic phase is followed by a presumably incommensurate collinear antiferromagnetic phase. Then a weak ferromagnetic canted A-type antiferromagnetic ordering is established at lower temperatures. For $0.2 \leq x \leq 0.4$, a different phase sequence is obtained. The canted A-type antiferromagnetic arrangement is no more stable, and instead a pure antiferromagnetic ordering is stabilized below $T_{lock} \approx 14 - 17$ K, with an improper ferroelectric character. From these results, a cycloid modulated spin arrangement at low temperatures is proposed, accordingly to the inverse Dzyaloshinskii–Moriya model. Anomalous temperature dependence of the dipolar relaxation energy and magnetization evidences for structural and magnetic changes occurring at $T^* \approx 22 - 28$ K for $0.1 \leq x \leq 0.4$.

The experimental results already available in this system point out for the importance of changing the balance between the exchange interactions underlying the macroscopic properties through the fine tuning of the microscopic structure, arising from Gd-substitution. In this respect, the study of the structural parameters, which can directly evidence the changes in the lattice, is absolutely necessary. Moreover, the emergence of electric polarization in modulated magnetic phases, as well as, the changes on the activation energy of the relaxation processes across the critical temperatures of the magnetic phase transitions evidence the existence of spin-lattice coupling in this system, which deserves a deeper study. In this work, we present an experimental study of the structure and lattice dynamics of the $Gd_{1-x}Y_xMnO_3$ system, with $x$ = 0.0, 0.1, 0.2, 0.3 and 0.4, at room temperature, through x-ray powder diffraction and Raman scattering, in order to unravel the effect of the lattice distortions induced by the controlled substitution of $Gd^{3+}$ with the smaller $Y^{3+}$ ions, namely the tilting of $MnO_6$ octahedra and the Jahn-Teller distortion. Moreover, the study of the structural distortions due to magnetic ordering through the temperature dependence of the Raman-active modes, as well as, the relation between phonon frequency and spin-spin correlation functions is also presented.

## 2. Experimental Details

High quality $Gd_{1-x}Y_xMnO_3$ ceramics, with $x$ = 0.0, 0.1, 0.2, 0.3, 0.4, 0.6, 0.8 and 1, were processed through the urea sol-gel combustion method, sintered at 1350 °C for 60 to 90 hours, and then quenched to room temperature. The rapid cooling is known to be efficient to guarantee the oxygen stoichiometry of the samples. Details of sample processing are available elsewhere.[5] The samples were characterized in terms of chemical, morphological and microstructure by using powder X-ray diffractometry, scanning electron microscopy, energy dispersive spectroscopy and X-ray photoemission spectroscopy techniques. The x-ray powder diffraction spectra of the $Gd_{1-}$

$_x$Y$_x$MnO$_3$ were recorded at room temperature using the X'Perto Pro PANalytical diffractometer, in the Bragg-Bentano geometry. The measurements were performed using the K$_{\alpha1}$ and K$_{\alpha2}$ doublet emitted by the Cu cathode, with wave numbers 1.540598 Å and 1.544426 Å, respectively. The diffractometer uses an X'Celerator detector and a secondary monocromator. The spectra were measured, in the 10° to the 70° 2θ range, with a step of 0.017° and an acquisition time of 100 s.step$^{-1}$. The calibration and alignment of the diffractometer were made by polycrystalline silica as external standard. The unpolarized Raman spectra were recorded using a T64000 Jobin-Yvon spectrometer, operating in triple subtractive mode, and equipped with a liquid nitrogen cooled charge-coupled device. The 514.5 nm linear polarized line of an Ar$^+$ laser was used for excitation. The Raman scattering studies were performed using polished pellets, with dimensions around 1 cm$^3$. The sample homogeneity was previously cheeked at room temperature. Thereafter, the Raman spectra of Gd$_{1-x}$Y$_x$MnO$_3$ were recorded in the pseudo-backscattering geometry at fixed temperatures in the temperature range 9 – 300 K. The samples were placed in a closed-cycle helium cryostat, with a temperature range from 9 K to 300 K, with a temperature stability of about 0.2 K. The temperature homogeneity in the samples was achieved with a cooper mask setup, estimated to differ by less than 1 K from the temperature measured with a silicon diode attached to the sample holder. The effect of the laser power on the Raman spectra was previously studied, in order to prevent the self-heating of the sample. In order to ensure reliable results, all of the Raman spectra were recorded at fixed holographic grating positions. Using 1800 lines·mm$^{-1}$ holographic gratings and taking into account the 3 mm × 640 mm focal length, the spectral resolution is 0.2 cm$^{-1}$. Identical conditions were maintained for all scattering measurements. The Raman spectra were fitted by using a sum of damped oscillators, according to Equation:

$$I(\omega,T) = [1 + n(\omega,T)] \sum_{j=1}^{N} A_{0j} \frac{\omega \Omega_{0J}^2 \Gamma_{0J}}{(\Omega_{0J}^2 - \omega^2)^2 + \omega^2 \Gamma_{0J}}, \qquad (1)$$

where $n(\omega,T)$ is the Bose-Einstein factor, $A_{0j}$ the strength, $\Omega_{0j}$ the wave number and $\Gamma_{0j}$ the damping coefficient of the $j$th oscillator.

### 3. Experimental results and discussion

#### a. Crystal structure at room temperature, Jahn-Teller and GdFeO$_3$-type distortions

Figure 1 shows representative x-ray powder diffraction patterns of the Gd$_{1-x}$Y$_x$MnO$_3$, with $x$ = 0.0, 0.2, 0.4, 0.6 and 1.0, recorded at room temperature, in the 10° to the 50° 2θ range.

The x-ray diffraction pattern of GdMnO$_3$ exhibits the typical profile observed for orthorhombic rare-earth manganites, with crystal structure described by the P*nma* space group.[6,7] For the compounds with *x* = 0.1 up to 0.4, the x-ray diffraction patterns are quite similar to the GdMnO$_3$ one, apart from a shift of the diffraction peaks toward higher 2θ values as the Y-content increases, as a consequence of the volume reduction. This is due to the substitution of Gd$^{3+}$ by the smaller Y$^{3+}$ ion, considered as a compressive hydrostatic pressure. The x-ray patterns for the range of compositions with 0 ≤ *x* ≤ 0.4, are consistent with the P*nma* space group.

The x-ray diffraction pattern of YMnO$_3$, on the top of Figure 1, presents the typical profile known for hexagonal rare-earth manganites, whit P*6$_3$cm* symmetry.[6] The *x* = 0.8 compound (not shown) evidences a very similar x-ray diffraction pattern to the YMnO$_3$; however, in this case, the diffraction peaks are shifted toward lower 2θ, as the volume increases when the Y-content decreases. For the *x* = 0.6 composition, the x-ray spectrum clearly shows Bragg peaks arising from both P*nma* and P*6$_3$cm* structures; for instance, the peaks at 2$\theta$ = 15.7°, 28.9°, 30.6° and 31.3°, assigned to the hexagonal P*6$_3$cm* phase, appear along with the reflected peaks assigned to the orthorhombic P*nma* phase. Figure 2 shows the molar percentage of orthorhombic and hexagonal phases as a function of Y-content, obtained from Rietveld analysis of the x-ray diffraction data. As the Y$^{3+}$ concentration increases from 0.4 towards 0.8, the molar percentage of P*nma* phase decreases while the molar percentage of P*6$_3$cm* phase increases. In particularly, for *x* = 0.6 almost 50% of each phase is found. In the following, we will focus in the range from *x* = 0.0 to 0.4, where the samples exhibit a pure orthorhombic P*nma* phase.

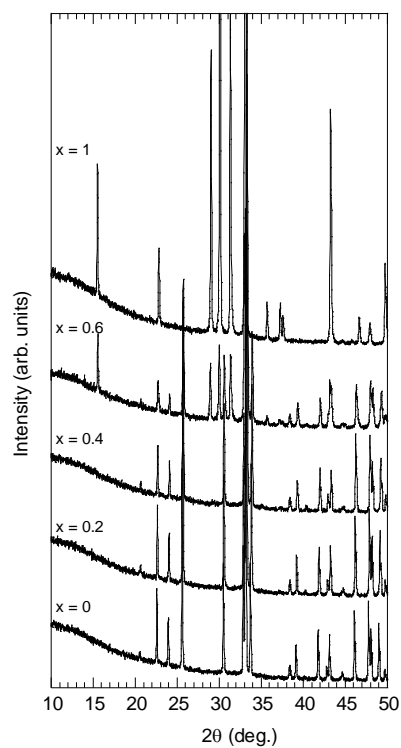

**Figure 1.** Representative x-ray powder diffraction patterns of $Gd_{1-x}Y_xMnO_3$, with $0 \leq x \leq 1$, recorded at room temperature.

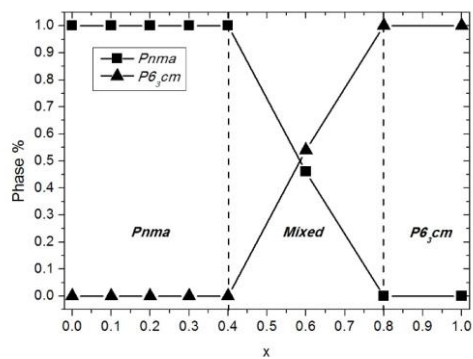

**Figure 2.** Molar percentage of orthorhombic and hexagonal phases for the $Gd_{1-x}Y_xMnO_3$ system.

From the Rietveld analysis of the x-ray patterns the Wyckoff positions and respective symmetry, and the atomic fractional coordinates were determined (see *Supplemental Material*). At room conditions, the unit cell of all studied compounds contains four formula units. Similarly to other orthorhombic rare-earth manganites, the crystal structure of $Gd_{1-x}Y_xMnO_3$, with $0.0 \leq x \leq 0.4$, consists on $MnO_6$ octahedra chains, developing along the *b*-direction and sharing one oxygen atom. The crystal structure of $Gd_{1-x}Y_xMnO_3$ exhibits the basic distortions observed in other orthorhombic rare-earth manganites, involving the [010] and [101] rotations of the $MnO_6$ octahedra, also known as tilting, and the cooperative Jahn-Teller distortion. The $Gd^{3+}/Y^{3+}$ ions (A-site cations), located at the center of the eight surrounding $MnO_6$ octahedra, are displaced from the high-symmetry position due to octahedra tilting and cooperative Jahn-Teller distortion and thus its coordination reduces from 12 to 8, but remaining in the mirror plane.

Figure 3 shows the Mn-O bond lengths and Mn-O-Mn bond angles as a function of Y-concentration. Three different Mn-O bond lengths were obtained, as it is expected from the cooperative Jahn-Teller distortion of the $MnO_6$ octahedron. No significant x-dependence of Mn-O bond lengths could be ascertained from the results displayed in Figure 3(a). This means that the Jahn-Teller distortion does not significantly change with the volume reduction due to the $Gd^{3+}$ substitution. Moreover, the [010] and [101] $MnO_6$ rotations are manifested by the two different values for the Mn-O-Mn bond angle. It is expected that the $Gd^{3+}$ substitution by the smaller $Y^{3+}$ ion induces reduction of the Mn-O-Mn bond angles, as it has been observed in the unsubstituted orthorhombic rare-earth manganites series (A = La to Dy) and Y- or Lu-substituted $EuMnO_3$. However, regarding the error bars of the Mn-O-Mn bond angle values observed in Figure 3(b), we cannot detect variations smaller than 2°.

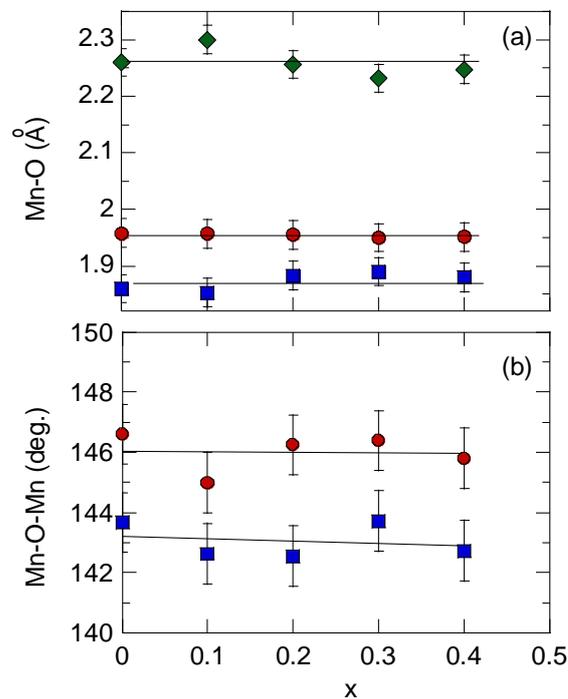

**Figure 3.** (a) Mn-O bond lengths (diamonds long Mn-$O_2$(l), circles Mn-O1, squares short Mn-O*2* (s)); and (b) Mn-O-Mn bond angles (circles Mn-O1-Mn; squares Mn-O*2*-Mn) as a function of the Y-concentration.

The octahedral rotations cause the long Mn-O2 bonds to align more closely with the $a$-axis than with the $c$-axis. The mean angles between the $a$- and $c$-axes and the short Mn-O2 bond are, respectively, 58.1° and 33.7°, and the long Mn-O2 bond are 30.7° and 60.6°. Due to these distortions, the four oxygen atoms belonging to the equatorial plane of the MnO$_6$ octahedra occupy general positions, while the Mn$^{3+}$ ion remains on the inversion center and the apical oxygen atoms remain in the mirror planes.

The lattice distortions induced by the Gd-substitution are better studied by analyzing the $x$-dependence of the lattice parameters. Figure 4 shows the pseudocubic lattice parameters and volume as a function of the Y-content, $x$ (definition of the pseudo-cubic lattice parameters are available in *Supplemental Material*).

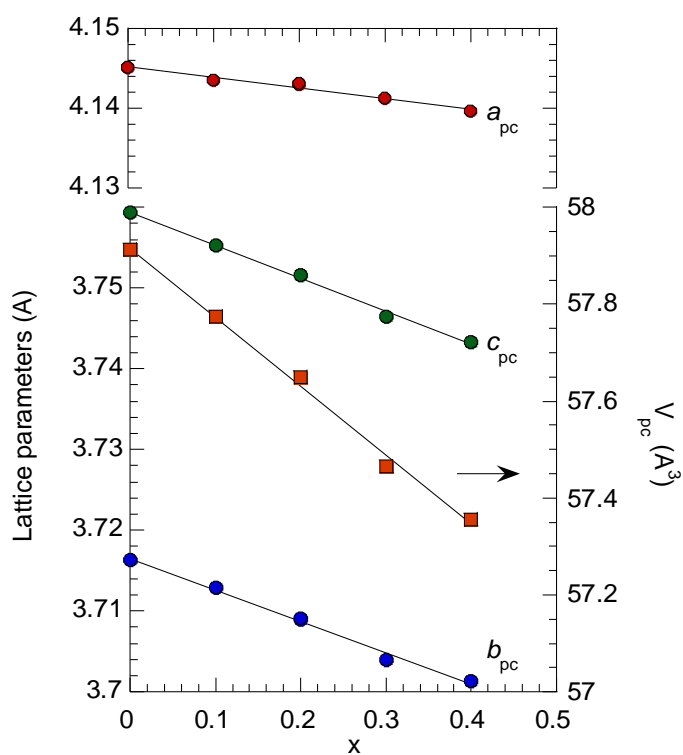

**Figure 4.** Pseudocubic lattice parameters (closed symbols) and volume (closed squares) as a function of Y-concentration. The error bars are smaller than the markers.

The pseudocubic lattice parameters well satisfy the relation $a_{pc} > c_{pc} > b_{pc}$, which has been typically found in perovskites presenting both octahedra tilting and Jahn-Teller distortion.[8,9] A linear decrease of the pseudocubic lattice parameters and volume with increasing $x$ is evident, as it is expected from the substitution of the Gd$^{3+}$ cation by the smaller Y$^{3+}$ one, regarded as a positive chemical pressure. Table 3 shows the absolute value of the relative change of the lattice parameters length, defined by equation:

$$\beta = \frac{1}{\ell(0)} \frac{d\ell}{dx} \qquad (2)$$

where $\ell(0)$ denotes the lattice parameter for $x$ = 0.0, and $d\ell/dx$ is the slope of the linear relations $\ell(x)$.

Table 3. Absolute value of the relative change of the lattice parameters length of $Gd_{1-x}Y_xMnO_3$.

| $\beta_a$ | $\beta_b$ | $\beta_c$ |
|---|---|---|
| 0.0032±0.0002 | 0.0105±0.0003 | 0.0109±0.0002 |

The $a$-axis is the hard direction, exhibiting the smallest relative change of the lattice parameter with x. Both $b$- and $c$-lattice parameters decrease with similar relative changes with $x$. The difference in the relative change of the lattice parameters lengths with $x$ points out for an anisotropic volume reduction of the unit cell, as it has been found in other solid solutions, like $Eu_{1-x}Y_xMnO_3$ and $Eu_{1-x}Lu_xMnO_3$.[5,10] The increase of the Y-amount increases the difference between the $a$- and $c$-lattice parameters, which reflects the decrease of the Mn-O-Mn bond angles, unbalancing the competition between the ferro and antiferromagnetic interactions, favoring the latter ones. The increase of the antiferromagnetic character of the $Gd_{1-x}Y_xMnO_3$ with increasing x, reflects the increase of the octahedral tilting.[11]

The effect of the Gd-substitution by $Y^{3+}$ on the volume reduction is different from the results obtained from high pressure studies on $GdMnO_3$, where the hard direction is the $b$-axis, and the $a$-axis being softer than the $c$-axis.[7] In Ref. [7], the effect of hydrostatic pressure was found to be more pronounced in the $ac$-plane, which contains the two Mn-O2 distances. Since the long and short Mn-O2 bonds have their main projections along the $a$ and $c$-axes respectively, the larger compressibility of the $a$-axis under hydrostatic pressure has been interpreted as a stronger reduction of the long Mn-O2 bond length as compared to the short one, and therefore a reduction of the Jahn-Teller distortion. In this work, we found that the volume reduction due to the Gd-substitution is smaller along the $a$-axis, pointing for a small variation of the long Mn-O2 bond length. This interpretation is consistent with the relative independence of both Mn-O2 bond lengths on the x-concentration, as it can be seen in Figure 3. So, we conclude that the substitution of $Gd^{3+}$ by $Y^{3+}$ does not induce significant additional distortions of the $MnO_6$ octahedra. Similar results has been found in $Eu_{1-x}A_xMnO_3$ system, with A = $Y^{3+}$ (0 ≤ $x$ ≤ 0.5) and

Lu$^{3+}$ (0 ≤ *x* ≤ 0.4).[5,10] Thus, the volume reduction already evidenced in this work can be mainly assigned to octahedral tilting, as it will be farther discussed in the next section.

### b. Lattice dynamics at room temperature and GdFeO$_3$-type distortion

The structural distortions referred to above are responsible for the Raman activation of several modes. As the Mn$^{3+}$ ion occupies a special position with symmetry $\bar{1}$, the Raman modes observed at room conditions do not involve motions of Mn$^{3+}$ ion. Factor group analysis provides the following decomposition corresponding to the 60 normal vibrations at the Γ-point of the Brillouin zone:

$$\Gamma_{acoustic} = B_{1u} + B_{2u} + B_{3u} \qquad (7)$$

$$\Gamma_{opt} = \left(7A_g + 5B_{1g} + 5B_{2g} + 5B_{3g}\right)_R + (9B_{1u} + 7B_{2u} + 9B_{3u})_{IR} + (8A_u)_{silent} \qquad (8)$$

Figure 6 shows the Raman spectra of Gd$_{1-x}$Y$_x$MnO$_3$, with *x* = 0.0, 0.1, 0.2, 0.3, and 0.4, recorded at room conditions, in the 200 – 800 cm$^{-1}$ spectral range. The Raman spectra of Gd$_{1-x}$Y$_x$MnO$_3$, with 0 ≤ *x* ≤ 0.4, exhibit the typical profile of the orthorhombic rare-earth manganites,[9,12] in good agreement with the structural data reported above. Due to the polycrystalline nature of the studied samples, the Raman spectra exhibit simultaneously all Raman-active modes A$_g$ and B$_{2g}$, which are the more intense bands.

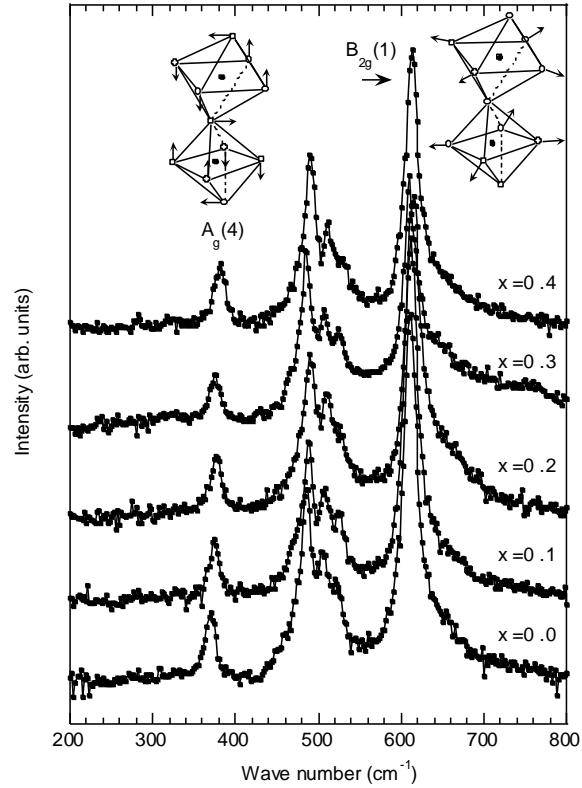

**Figure 6.** Unpolarized Raman spectra of $Gd_{1-x}Y_xMnO_3$, recorded at room temperature, for $x$ = 0.0, 0.1, 0.2, 0.3, and 0.4. The out-of-phase $MnO_6$ rotation mode ($A_g(4)$) and the in-plane O2 symmetric stretching mode ($B_{2g}(1)$) are labelled in this Figure. The sketches show the atomic motions involved in these modes.

We follow the well stablished mode assignment published for orthorhombic rare-earth manganites,[12–14] and we use the notation defined in Ref. [12]. By fitting Equation (1) to the experimental spectra, we have determined the wave number of the observed Raman bands, which is depicted as a function of the Y-concentration in Figure 7(a). As the Y-concentration increases, the Raman bands shift towards higher wave numbers, due to volume reduction, although with different slopes. Table 2 presents the slope of the linear relation between wave number and Y-concentration.

**Table 2.** Slope of the linear relation between the wave number of the active Raman bands and the Y-concentration.

| $\Delta\omega/\Delta x$ (cm$^{-1}$) | | | | | |
|---|---|---|---|---|---|
| $Ag(4)$ | $B2g(3)$ | $Ag$ | $Ag$ | $B2g(2)$ | $B2g(1)$ |
| 29±2 | 19±6 | 12±1 | 14±1 | 10±3 | 3±2 |

In the following, we focus our attention on the Raman bands whose activation is associated with the main structural deformations allowing for symmetry reduction. We have labelled these bands in Figure 6, accompanied by a sketch of the modes, showing the atomic displacements associated with each mode. These modes are the in-plane O2 symmetric stretching mode, of symmetry $B_{2g}$, activated by the Jahn-Teller distortion, and the out-of-phase $MnO_6$ rotation mode, of symmetry $A_g$, activated by the [101] rotations. The in-plane O2 symmetric stretching mode involves the stretching vibrations of the equatorial plane O2 atoms, whose wave number is determined by the mean Mn-O2 distance,[12] and thus, this mode probes the relative octahedra distortion in the Mn-O2 plane, resulting from the oxygen displacements. As the Mn-O2 bond are involved in the superexchange pathway in rare-earth manganites, this mode affects electronic orbital overlap. From Table 2, we conclude that the in-plane O2 stretching mode is quite independent on the volume reduction, evidencing tiny variations of the Mn-O2 bond lengths with increasing x, as it was discussed by the analysis of the x-dependence of the Mn-O2 bond lengths. The out-of-phase $MnO_6$ rotation mode is known to scale with the [101] tilt angle.[12] Figure 7(b) shows the wave number of the out-of-phase $MnO_6$ rotation mode for unsubstituted rare-earth manganites (data taken from Ref. [12]) and Y-substituted $GdMnO_3$. The same linear relation between Mn-O1-Mn bond angle and the wave number of out-of-phase $MnO_6$ rotation mode is clearly ascertain for both unsubstituted rare-earth manganites and Y-substituted $GdMnO_3$, meaning that the compression mechanism induced by the chemical pressure acts in similar way in the tilting of the $MnO_6$ octahedra chains, and thus, change the balance between the competitive magnetic interactions. This result corroborates the possibility to tune the lattice deformations through chemical substitution in the A-site towards magnetoelectricity in rare-earth manganites.

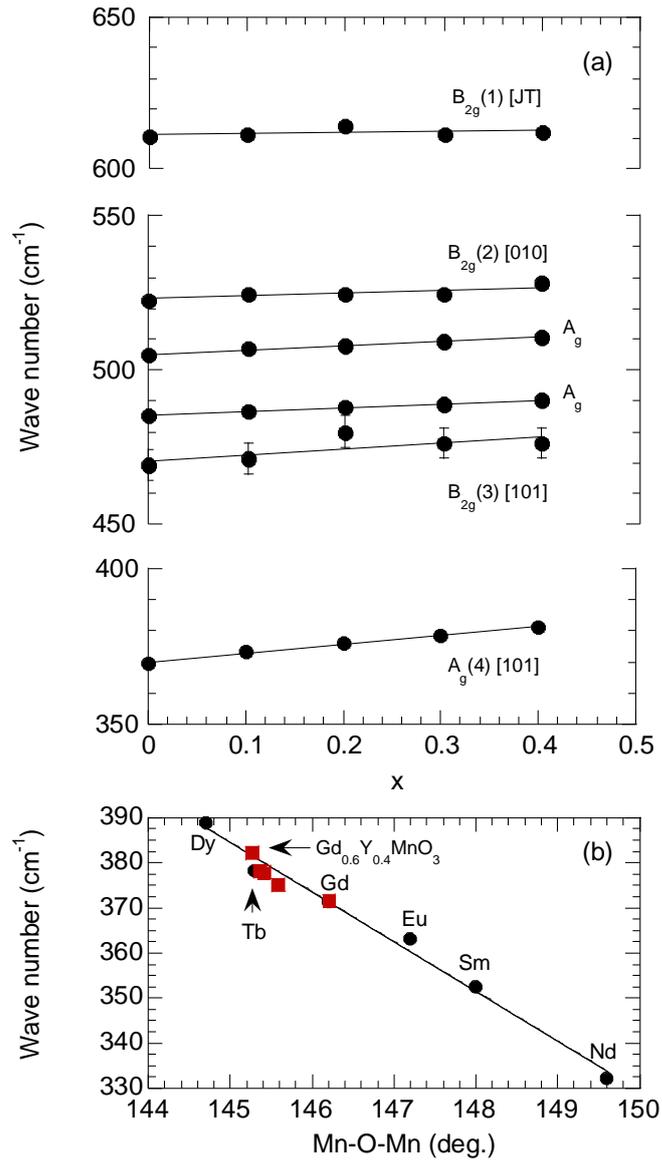

**Figure 7.** (a) x-dependence of the wave number of the Raman bands, and mode assignment according to Ref. [12]. The basic distortion responsible for the mode activation is also presented. $A_g(4)$: out-of-phase $MnO_6$ [101] rotation, $B_{2g}(3)$ out-of-phase $MnO_6$ [101] rotation, $B_{2g}(2)$ in-phase $O_2$ scissorslike [010] rotation, $B_{2g}(1)$: in-plane $O_2$ stretching Jahn-Teller, and $A_g$ mixed modes. The solid lines are guides for the eyes. (b) Wave number of the out-of-phase $MnO_6$ rotation mode as a function of the Mn-O-Mn bond angle of unsubstituted rare-earth manganites (close circles) and $Gd_{1-x}Y_xMnO_3$ (close squares). Data for unsubstituted manganites were taken from Ref. [12].

The larger change rate with *x* is found for the out-of-phase rotation $A_g(4)$ mode (see Table 2) which enables us to estimate the changes on the Mn-O1-Mn bond angle with *x*. Considering that this mode has the same linear dependence with the tilt angle of 23.5 cm$^{-1}$/deg, as demonstrated by Iliev et al,[12] we can expect variations of $\frac{\Delta\theta}{\Delta x} = 1.3°$. So, the largest increase of the tilt angle induced by the volume reduction in $Gd_{1-x}Y_xMnO_3$, with $x$ up to 0.4, is about 0.5°. This result confirms the increase of the $GdFeO_3$-type distortion in $Gd_{1-x}Y_xMnO_3$ as $x$ increases, albeit being too small to be undetected through XRD analysis.

### c. Lattice dynamics at low temperatures and spin-phonon coupling

Figure 8 shows the unpolarized Raman spectra recorded at several fixed temperatures in the 9 K to 300 K temperature range, for the compounds with $x$ = 0.0, 0.2 and 0.4, as representative examples of the temperature evolution of the Raman signal.

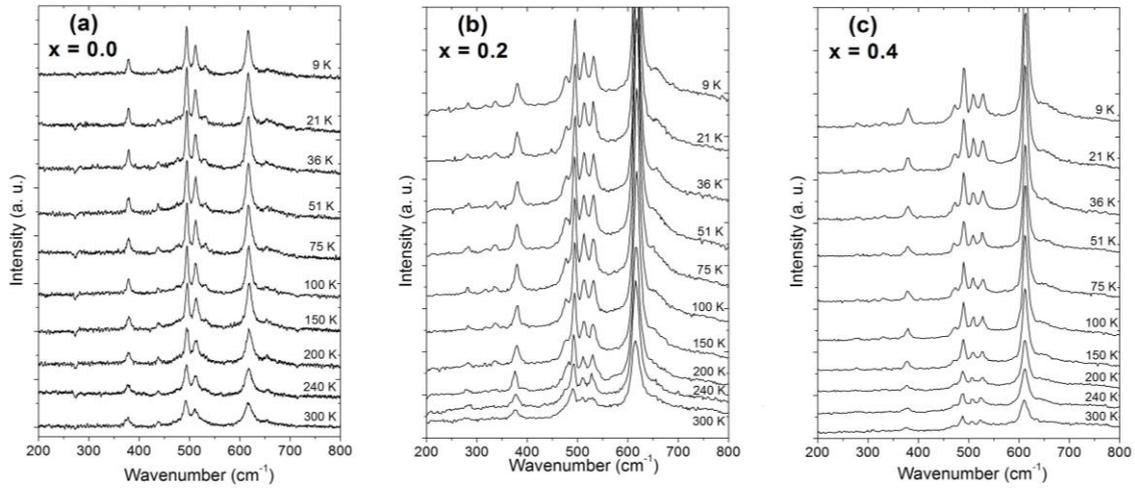

**Figure 8.** Raman spectra of $Gd_{1-x}Y_xMnO_3$, with $x$ = 0.0, 0.2, and 0.4, recorded at several fixed temperatures in the 9 K to 300 K temperature range.

As temperature decreases, the Raman bands become narrower and better resolved, due to the decrease of disorder arising from thermal motions. No new Raman bands were detected at low temperatures, even in the magnetic phases which allow for ferroelectricity.[4] The absence of new well-defined activated Raman bands and the prevalence of the high-temperature spectra profile point out for the weak polar character of these compounds associated with small atomic dislocations, and corroborate the improper nature of the ferroelectric phases, as is reported in Refs. [5,10,15]. In fact, as the electric polarization of these compounds is very low (dozen of pC/cm²)[5,10,15], the intensity of the new Raman bands, which is proportional to the oscillator strengths of the infrared active modes, is very small and could not be detected with the available experimental conditions.

In order to calculate the temperature anomalous behavior of the frequency due to the magnetic transitions, we have described the purely anharmonic temperature dependence of the frequency by the model:[16]

$$\omega(T) = \omega(0) + C\left(1 - \frac{2}{e^x - 1}\right), \tag{10}$$

where $C$ is a model constant, $x$ is given by $x = \frac{\hbar \omega_0}{2k_B T}$, $\hbar$ is the reduced Planck constant, $k_B$ is the Boltzmann constant and $T$ is temperature. Figures 9 shows the temperature dependence of the wave number of the in-phase O2 stretching mode, respectively, for the different compositions, along with the curve calculated from the best fit of Equation 10 to the experimental data above 100 K, and extrapolated to lower temperatures.

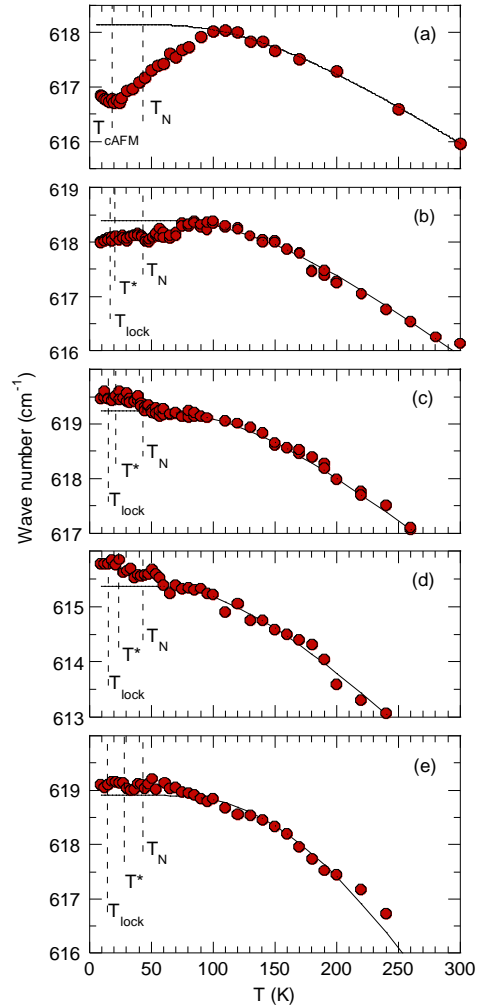

**Figure 10.** Wave number of the in-phase O2 stretching modes as a function of temperature for GdMnO$_3$ (a), Gd$_{0.9}$Y$_{0.1}$MnO$_3$ (b), Gd$_{0.8}$Y$_{0.2}$MnO$_3$ (c), Gd$_{0.7}$Y$_{0.3}$MnO$_3$ (d), and Gd$_{0.6}$Y$_{0.4}$MnO$_3$ (e). The solid lines were determined by fitting Eq. 10 to the experimental data above 100K, which was extrapolated to lower temperatures. The dashed lines mark the magnetic phase transitions, according to Ref. [4].

A downward deviation of the wave number of the analyzed modes from the extrapolated anharmonic behavior is evident below 100 K and ≈ 75 K, for GdMnO$_3$ and Gd$_{0.9}$Y$_{0.1}$MnO$_3$, respectively. This deviation starts to be observed well above T$_N$ = 42 K, and reflects the local ordering of the Mn$^{3+}$ magnetic momenta. This is a sign of precursor effects of the magnetic phase transitions, as it is the case of EuMnO$_3$ compound and its Y-substituted parents.[17] It is worth to stress that the deviation is more pronounced in the case of GdMnO$_3$ than in the Gd$_{0.9}$Y$_{0.1}$MnO$_3$. In contrast, for the compositions with *x* = 0.2 to 0.4 an upward deviation, starting just above T$_N$, is clearly observed.

The transition from the sinusoidal incommensurate antiferromagnetic to the canted A-type antiferromagnetic phase, taking place in GdMnO$_3$ at T$_{cAFM}$ = 20 K, is well marked by anomalous behavior of the temperature dependence of the wave number, which is reflected by a change of slope of the $\omega(T)$ curve. For the remaining compositions, the experimental resolution prevents us from observing significant anomalies of the $\omega(T)$ curve at the critical temperatures, already determined from macroscopic characterization of this system.[4]

The deviation of the temperature dependence of the wave number relatively to the anharmonic temperature behavior, extrapolated below 100K, evidences the existence of the spin-phonon coupling in this system. Within the scope the spin-phonon coupling model, the wave number deviation of the phonon as a function of temperature is determined by the spin-spin correlation function.[18] When competitive ferro and antiferromagnetic interactions exist, and assuming that the spin-spin correlation function of both the nearest neighbors and the next-nearest neighbors for the Mn$^{3+}$ spins has the same temperature dependence, we obtain:[18]

$$\omega(T) - \omega_0 \approx (R_{AFM} - R_{FM}) <S_i \cdot S_j>, \qquad (11)$$

where $R_{FM}$ and $R_{AFM}$ are spin dependent force constants of the lattice vibrations, defined as the squared derivatives of the ferro and antiferromagnetic exchange integrals, respectively, with respect to the normal coordinate.[18] According to Equation 11, deviations of the Raman modes frequencies, relatively to the normal temperature behavior, are expected in the magnetic ordered phases, where the balance of competitive ferromagnetic and antiferromagnetic interactions depends on the Y-concentration. For *x* = 0.0 and 0.1, a negative deviation is observed, decreasing its magnitude as the *x* value increases up to 0.1. Although GdMnO$_3$ is an antiferromagnetic material below T$_N$, the canted nature of the spin arrangement below T$_{cAFM}$ yields a weak-ferromagnetic character. In the case of Gd$_{0.9}$Y$_{0.1}$MnO$_3$, a non-vanishing difference between the temperature dependence of the magnetization measured in zero-field cooling and field cooling conditions suggests a weaker ferromagnetic character. The downward shift of the wave number as a function of temperature still points to the relative importance of the ferromagnetic interactions. As $Gd^{3+}$ ions are being farther substituted by $Y^{3+}$ ones, an increase of the relative importance of the antiferromagnetic interactions against the ferromagnetic ones is observed, and the deviations of the Raman modes shift from negative to positive values, as it is clearly ascertained for *x* = 0.2 and 0.4. This result is in good agreement with the antiferromagnetic character of these compounds.[4]

Once the existence of spin-phonon coupling is established, it is important to determine its strength. Quantitatively, the spin-spin correlation function can be directly calculated from the integral of the magnetic contribution to the specific heat $C_m$:

$$\int_{200}^{10} C_m dT = 6J <S_i \cdot S_j>, \qquad (12)$$

where *6* stands for the number of $Mn^{3+}$ nearest-neighbors, and *J* for the effective magnetic exchange constant. The specific heat data were taken from our previous work.[4] We have interpolated the data resulting from the temperature integration of the $C_m$(T), in order to obtain the (<$S_i.S_j$>, ω). Points referred to the same temperature as the wave number of the Raman spectra. In this analysis, we have chosen the in-plane O2 stretching mode to be scaled by the spin-spin correlation function, as this mode, affecting the overlapping between 3*d*-$M^{3+}$ and 2*p*-$O^{2-}$ electronic orbitals, is sensitive to the exchange magnetic interactions. Figure 11 presents the plot of the wave number of the in-plane O2 stretching mode against the spin-spin correlation function, for all studied compounds, in the temperature range where monotonous temperature behavior of the wave number is observed. A linear behavior of the in-plane O2 stretching mode wave number as a function of the spin-spin correlation function is evident, wherein the slope depends on the Y-composition. In the case of $GdMnO_3$, a change of slope of the linear relation between wave number and spin-spin correlation function is observed at $T_{cAFM}$, revealing that a change of the spin-phonon coupling occurs at that temperature. For the compositions with *x* from 0.1 up to 0.4, no anomalous behavior could be observed at the critical temperatures.

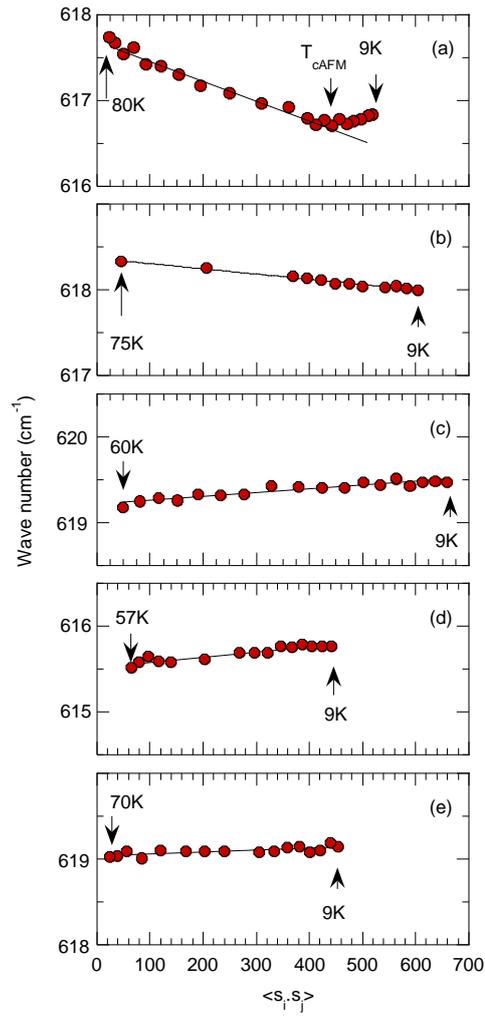

**Figure 12.** Wave number of the in-plane O2 stretching mode as a function of the spin-spin correlation function, for GdMnO$_3$ (a), Gd$_{0.9}$Y$_{0.1}$MnO$_3$ (b), Gd$_{0.8}$Y$_{0.2}$MnO$_3$ (c), Gd$_{0.7}$Y$_{0.3}$MnO$_3$ (d), and Gd$_{0.6}$Y$_{0.4}$MnO$_3$ (e).

The ratio λ/J between the spin-phonon coupling parameter and the effective magnetic exchange constant per spin are presented in Table 4. The value of λ/J for GdMnO$_3$ is the largest one, taking the value 0.015±0.002 cm$^{-1}$.eV$^{-1}$. For the other compositions, λ/J ranges between 0.005 – 0.002 cm$^{-1}$eV$^{-1}$, about 3 to 5 times less than the value for GdMnO$_3$.

| Composition | λ/J (cm$^{-1}$.eV$^{-1}$) |
|---|---|
| x = 0.0 | 0.015±0.002 |
| x = 0.1 | 0.003±0.001 |
| x = 0.2 | 0.003±0.001 |
| x = 0.3 | 0.005±0.001 |
| x = 0.4 | 0.002±0.001 |

Table 4. Ratio between the spin-phonon coupling parameter and the magnetic exchange constant.

## 4. Conclusions

In this work, we report a systematic study concerning the structure and lattice dynamics at room temperature, and the phonon behavior at low temperatures in orthorhombic $Gd_{1-x}Y_xMnO_3$ manganites, with $x$ ranging from 0.0 up to 0.4.

The structural data obtained from both x-ray diffraction and Raman scattering experiments are consistent with the orthorhombic P*nma* symmetry, revealing lattice deformations associated with both the cooperative Jahn-Teller distortion and octahedra tilting. While the Jahn-Teller distortion is apparently x-independent, an increase of Mn-O1-Mn bond angle of about 0.5° could be ascertain from the x-dependence of the wave number of the out-of-phase $MnO_6$ rotations, when $x$ ranges from 0 to 0.4. The increase of the octahedra tilting unbalances the competition between ferro and antiferromagnetic exchange interactions, yielding an increase of the antiferromagnetic character with increasing $x$.

Clear evidences for spin-phonon coupling was ascertained from Raman results. The downward shift of the wave number of the in-plane O2 stretching mode, relatively to the anharmonic temperature behavior, observed below 100 K (for $x$ = 0) and 70 K (for $x$ = 0.1) points to the relative importance of the ferromagnetic interactions in these compositions. On the contrary, the upward shift of the wave number of the same mode, ascertained for the remaining compositions, reveals the reinforcement of the antiferromagnetic interactions against the ferromagnetic ones. This result corroborates the progressive disappearance of the weak-ferromagnetic character observed for the compositions $0 \leq x \leq 0.1$, and the emergence of pure antiferromagnetic ordering at low temperatures, for compositions with $0.2 \leq x \leq 0.4$. The emergence of such a low temperature antiferromagnetic phase is compatible with ferroelectricity, according to the Dzyaloshinskii-Moriya model.

From the Raman results, the observed deviation of the frequency from the anharmonic behavior, above the magnetic phase transition, is associated with the spin-phonon coupling. If we consider that manganites are well-known for their often peculiar local structure in terms of electronic and magnetic properties, it is plausible to associate this type of behavior with local magnetic fluctuations.

We have correlated the deviation of the wave number of the in-plane O2 stretching mode relative to the extrapolated anharmonic temperature dependence, with the spin-spin correlation function, calculated from the analysis of the magnetic contribution to the specific heat. From the linear relation obtained, the ratio $\lambda/J$ between the spin-phonon coupling parameter and the effective magnetic exchange constant per spin was determined. This ratio is higher for the GdMnO$_3$, and decreases down to 20% for the remaining compositions.

Finally, we would like to stress that the anomalies of the temperature dependence of the wave number of the in-plane O2 stretching mode, only involving the motion of the O2 atoms, also corroborates the active role of oxygen displacements in the stabilization of both polar and magnetic phases.

**Acknowledgements**

The Authors are very grateful to M. Sá, J. M. B. Oliveira J. M. Machado da Silva for the measurements of the specific heat of the samples. This work was supported by Fundação para a Ciência e Tecnologia, through the Project PTDC/FIS-NAN/0533/2012 and by QREN, through the Project Norte-070124-FEDER-000070 Nanomateriais Multifuncionais.

# Structural properties and spin-phonon coupling in orthorhombic Y-doped GdMnO$_3$

**R. Vilarinho[1], A. Almeida[1], P. Tavares[2], J. Agostinho Moreira[1]**

1-IFIMUP and IN-Institute of Nanoscience and Nanotechnology, Departamento de Física e Astronomia da Faculdade de Ciências, Universidade do Porto, Rua do Campo Alegre, 687, 4169-007 Porto, Portugal.

2-Centro de Química - Vila Real, Departamento de Química. Universidade de Trás-os-Montes e Alto Douro, 5000-801 Vila Real, Portugal.

**Table 1**. Wyckoff positions and site symmetry in orthorhombic Gd$_{1-x}$Y$_x$MnO$_3$.

| Atom | Wyckoff position | symmetry |
| --- | --- | --- |
| $A$ | 4c | $m$ |
| Mn | 4b | $\bar{1}$ |
| O1 | 4c | $m$ |
| O2 | 8d | 1 |

**Table 2.** Atomic fractional coordinates obtained from the Rietveld refinement of the x-ray patterns.

| Composition | Atom | x | y | z |
| --- | --- | --- | --- | --- |
| 0.0 | Mn | 0.0000 | 0.0000 | 0.50000 |
|  | $A$ = Gd,Y | 0.080955 | 0.25000 | -0.018610 |
|  | O1 | 0.46800 | 0.25000 | 0.10883 |
|  | O2 | 0.17074 | 0.54229 | 0.20932 |
| 0.1 | Mn | 0.0000 | 0.0000 | 0.50000 |
|  | $A$ = Gd,Y | 0.080700 | 0.25000 | -0.017900 |
|  | O1 | 0.46360 | 0.25000 | 0.10970 |
|  | O2 | 0.16280 | 0.55070 | 0.20940 |
| 0.2 | Mn | 0.0000 | 0.0000 | 0.50000 |
|  | $A$ = Gd,Y | 0.081700 | 0.25000 | -0.018400 |
|  | O1 | 0.47130 | 0.25000 | 0.11000 |
|  | O2 | 0.16720 | 0.54780 | 0.20960 |
| 0.3 | Mn | 0.0000 | 0.0000 | 0.50000 |
|  | $A$ = Gd,Y | 0.081370 | 0.25000 | -0.016700 |
|  | O1 | 0.46180 | 0.25000 | 0.10610 |
|  | O2 | 0.17660 | 0.55340 | 0.20910 |
| 0.4 | Mn | 0.0000 | 0.0000 | 0.50000 |
|  | $A$ = Gd,Y | 0.082600 | 0.25000 | -0.016500 |
|  | O1 | 0.45980 | 0.25000 | 0.10800 |
|  | O2 | 0.17190 | 0.55210 | 0.21130 |

**Table 3.** Definition of the pseudo-cubic lattice parameters and components of the spontaneous strain

| Pseudo-cubic lattice constants for orthorhombic symmetry |
|---|
| $a_{pc} = \dfrac{a}{\sqrt{2}}$ |
| $b_{pc} = \dfrac{b}{2}$ |
| $c_{pc} = \dfrac{c}{\sqrt{2}}$ |